\documentstyle[12pt]{article}
\newcommand{\beq}{\begin{equation}}
\newcommand{\eeq}{\end{equation}}
\newcommand{\bea}{\begin{eqnarray}}
\newcommand{\eea}{\end{eqnarray}}

\begin{document}

\title{Regularized overlap and the chiral determinant}
\author{\rule{0cm}{1.cm} C.D. Fosco \thanks{%
Member of CONICET, Argentina.} ~and~ R.C. Trinchero \thanks{%
Member of Instituto Balseiro and CONICET , Argentina.} \\
{\normalsize {\it Centro At\'omico Bariloche}} {\normalsize {\it 8400 S.C.
de Bariloche, Argentina}}}
\maketitle

\begin{abstract}
We study the relationship between the continuum overlap and its
corresponding chiral determinant, showing that the former amounts to an
unregularised version of the latter. We then construct a regularised
continuum overlap, and consider the chiral anomalies that follow therefrom.
The relation between these anomalies and the ones derived from the formal
(i.e., unregularised) overlap is elucidated.
\end{abstract}

\section{Introduction.}

Chiral Gauge Theories~\cite{ball}, namely, the ones where the left and right
components of a fermionic field couple asymmetrically to a gauge field are
of fundamental importance in High Energy Physics.

Their study is of relevance, for example, to understand the dynamics of the
Standard Model, where their characteristic properties have important
consequences, such as the anomaly cancelation conditions.

To find a proper definition of the effective action for a fermionic field in
the presence of an external gauge field is usually the first step towards a
full quantization of the theory. This step is equivalent to the definition
(if possible beyond the perturbative level) of a chiral determinant. Many
attempts have been done to achieve this goal, and in recent years some new
proposals emerged, which, in spite of being at first sight different, are
related. They share the property of introducing an infinite number of
fermionic fields for each point in spacetime. This appears as an extra
dimension in Kaplan's formulation~\cite{kap}, as an infinite number of
Pauli-Villars fields in Slavnov's proposal~\cite{slav1,slav2}, and as a
continuum of fermionic degrees of freedom living inside each lattice cell in
t'Hooft approach~\cite{toft}.

The overlap formulation~\cite{nar} (inspired in an earlier idea of Kaplan~%
\cite{kap}) seems to correctly define a fermionic chiral determinant when
constructed on a lattice. In this letter we study the overlap in the
continuum, showing explicitly that it is equivalent to a {\em non-regularised%
} version of its corresponding chiral determinant. For the particular case
of the {\em modulus\/} of the chiral determinant in two spacetime dimensions
it yields a finite result.

It has been shown (\cite{daemi1, daemi2}) that the formal (i.e.,
unregularized) overlap in the continuum yields anomalies when its gauge
variation is calculated in a perturbative approach. A similar calculation
has been done on the lattice (\cite{daemi3,daemi4}), to show that the
continuum results are recovered when the lattice cutoff tends to zero.

We show that the anomalies derived from the unregularised overlap may lead
to ambiguous results when calculating the covariant divergence of the chiral current in terms of
the full fermion propagator in the presence of an external gauge field, and
then define a properly regularised continuum overlap, which leads to
consistent anomalies.


\section{Definitions and conventions.}

We shall consider the overlap associated to a chiral Dirac operator $\not
\!\! D_L =\not \!\! D P_L$, where $\not \!\!D \,=\,\gamma_\mu (\partial_\mu
\,+\, A_\mu )$, $P_L = \displaystyle{\frac{1+\gamma_5}{2}}$ is the left
chirality projector, and Dirac's matrices are chosen according to the
representation:
\begin{equation}
\label{dirac}\gamma_\mu \;=\; \left(
\begin{array}{cc}
0 & \sigma^\dagger_\mu \\
\sigma_\mu & 0
\end{array}
\right) \;\; , \;\; \gamma_5 \;=\;\left(
\begin{array}{cc}
1 & 0 \\
0 & -1
\end{array}
\right) \;\;,\;\; \sigma_\mu \,{\stackrel{d=4}{=}}\, ({\vec \sigma},1)\;,\;
\sigma_\mu \,{\stackrel{d=2}{=}}\, (1,i) \;,
\end{equation}
where ${\vec \sigma}$ denote the three familiar Pauli's matrices. $A_\mu (x)$
is a (anti-hermitian) gauge connection for the gauge group ${\cal G}$. It
may be written as $A_\mu (x)\,=\, A_\mu^a (x) \tau_a$, with real components $%
A_\mu^a (x)$ , $\tau_a^\dagger = - \tau_a$ , and $[\tau_a , \tau_b] \,=\,
f_{abc} \, \tau_c$. The overlap ${\cal O}$ corresponding to $\not \!\!\! D_L$
is defined, following Narayanan and Neuberger, by
\begin{equation}
\label{over}{\cal O} \;=\; \displaystyle{\frac{\langle + | A + \rangle}{%
|\langle + | A + \rangle|}} \;\;\; \displaystyle{\frac{\langle A + | A -
\rangle}{\langle + | - \rangle}} \;\;\; \displaystyle{\frac{\langle A - | -
\rangle}{|\langle A - | - \rangle|}}\;,
\end{equation}
where $|A \pm \rangle$ are the Dirac vacua corresponding to the Hamiltonians
\begin{equation}
\label{fqh}{\cal H}_\pm \;=\; \gamma_5 ( \not \!\! D \pm M ) \;=\; \left(
\begin{array}{cc}
\pm M & - C^\dagger \\
- C & \mp M
\end{array}
\right) \; ,
\end{equation}
respectively, and $|\pm \rangle = |A \pm \rangle |_{A = 0}$. The operator $C$
is defined by $C = \sigma_\mu (\partial_\mu + A_\mu)$. Scalar products
between the different Dirac vacua appearing in (\ref{over}) may be put in
terms of Slater determinants built from negative energy eigenspinors of the
corresponding first-quantized Hamiltonians~\cite{nar}. These negative energy
eigenspinors satisfy
\begin{equation}
\label{eigen}{\cal H}_\pm \; |v_\pm (\lambda)\rangle \;=\; - \, \omega
(\lambda) \; |v_\pm (\lambda)\rangle \;\qquad \omega (\lambda)>0,
\end{equation}
with the index $\lambda$ labeling the eigenstates~\footnote{%
We use the same ket symbol $| \; \rangle$ to denote second-quantization
vacua and first-quantization eigenspinors, since no confussion should arise.}%
.

The energy $\omega (\lambda)$ is independent of the sign of $M$, and
normalized negative energy eigenstates can be constructed as follows~%
\footnote{%
We shall assume at this stage that there are no zero modes either for $C$ or
$C^\dagger$. Had zero modes been present, the chiral determinant would have
been equal to zero.}
\begin{equation}
\label{normeig}|v_\pm (\lambda)\rangle \,=\, \sqrt{\frac{\omega (\lambda)
\mp M}{2 \, \omega (\lambda) }} \, \left(
\begin{array}{c}
|v_1 (\lambda)\rangle \\
\\
\displaystyle{\frac{C}{\omega (\lambda) \mp M}} \, |v_1 (\lambda) \rangle
\label{veq}
\end{array}
\right)\;,
\end{equation}
where $|v_1 (\lambda)\rangle$ are orthonormal eigenfunctions of the positive
hermitian operator $C^\dagger C$, satisfying
\begin{equation}
\label{v1eq}C^\dagger C \, |v_1 (\lambda)\rangle \;=\; ( \omega^2 (\lambda)
- M^2) \; |v_1 (\lambda) \rangle \;.
\end{equation}
The number of components of $|v_1 (\lambda) \rangle$ is half the one of $%
|v_\pm (\lambda)\rangle$. Of course, the eigenfunctions $|v_\pm
(\lambda)\rangle$ are not uniquely determined, since their phases may be
changed without affecting the normalisation. It will turn out, however, that
(\ref{veq}) is a convenient choice since it makes the factor $\displaystyle{%
\frac{{\langle A + | A - \rangle}}{{\langle + | - \rangle}}}$ in (\ref{over}%
) real, while the remaining two factors are pure phases. This makes the
identification of the modulus and phase of ${\cal O}$ easier.

As a straightforward calculation shows, the relevant scalar products in (\ref
{over}) may be put in terms of the above-defined eigenspinors as
\begin{equation}
\label{sprd}\langle A + | A - \rangle \;=\; \det_{\lambda ,
\lambda^{\prime}} \, \langle v_+ (\lambda) | v_- (\lambda^{\prime}) \rangle
\;\;,\;\; \langle \pm | A \pm \rangle \;=\; \det_{\lambda ,
\lambda^{\prime}} \, \langle v^0_\pm (\lambda) | v_\pm (\lambda^{\prime})
\rangle\;,
\end{equation}
where $|v^0_\pm (\lambda)\rangle$ denotes the negative-energy eigenfunction $%
|v_\pm (\lambda)\rangle$ for $A_\mu = 0$.

According to the overlap formalism, the effective action following from $%
{\cal O}$ should coincide, when $M \to \infty$, with the normalized chiral
Dirac operator effective action, i.e.,
\begin{equation}
\label{oc}\lim_{M \to \infty} \Gamma_{{\cal O}} (A,M)\,\equiv\,- \lim_{M \to
\infty} \log {\cal O} (A,M) \,=\, -\log \det ( {\not \!\partial}_L^{-1}{\not
\!\! D}_L )\,\equiv\, \Gamma_L (A) \;.
\end{equation}
It is convenient to split $\Gamma_{{\cal O}}$ into its real and imaginary
parts, since they enjoy very different properties (akin to the ones of the
real and imaginary parts of $\Gamma_L$). The real part of $\Gamma_{{\cal O}}$
comes from the factor $\displaystyle{\frac{\langle A + | A - \rangle}{%
\langle + | - \rangle}}$. By using (\ref{sprd}), (\ref{v1eq}) and (\ref
{normeig}) it is possible to write $\langle A + | A - \rangle$ as the
determinant of an operator acting on the Hilbert space generated by the set $%
\{ |v_1 (\lambda)\rangle \}$ of eigenfunctions of $C^\dagger C$
\begin{equation}
\langle A + | A - \rangle \;=\; \det \left( \frac{C^\dagger C}{C^\dagger C +
M^2}\right)^{1/2} \;.
\end{equation}
Then
$$
{\rm Re} \; \Gamma_{{\cal O}}\;=\; - {\rm Re} \,\log \left[ \frac{\langle A
+ | A - \rangle}{\langle + | - \rangle} \right]
$$
\begin{equation}
\label{real}=\;- \displaystyle{\frac{1}{2}} {\rm Tr} \log \left[ \frac{%
C^\dagger C}{C^\dagger C +M^2}\right] \;+\;\displaystyle{\frac{1}{2}} {\rm Tr%
} \log \left[ \frac{C_0^\dagger C_0}{C_0^\dagger C_0 +M^2} \right] \;\;,
\end{equation}
where, of course, the functional trace is also over the Hilbert space
generated by $\{ |v_1 (\lambda)\rangle \}$. The {\em formal\/} $M \to \infty$
limit of (\ref{real}) is
\begin{equation}
{\rm Re} \Gamma_{{\cal O}} \, \to \, - \frac{1}{2} {\rm Tr} \log \left(
\frac{C^\dagger C}{M^2} \right) \,+ \,\frac{1}{2} {\rm Tr} \log \left( \frac{%
C_0^\dagger C_0}{M^2} \right) \;,
\end{equation}
which may also be put in terms of Dirac operators as
\begin{equation}
{\rm Re} \Gamma_{{\cal O}} \, \to \, - \frac{1}{2} {\rm Tr} \log \left( \not
\!\! D^\dagger_L \not \!\! D_L \right) \,-\,\frac{1}{2} {\rm Tr} \log \left(
\not \!\! \partial^\dagger_L \not \!\! \partial_L \right) \;.
\end{equation}
which is, of course, unregularized. To see the role played by $M$, we note
that, keeping it finite, ${\rm Re} \, \Gamma_{{\cal O}}$ may be written as
$$
{\rm Re} \, \Gamma_{{\cal O}}(A,M) \, = \, - \frac{1}{2} \sum_{s=0}^1 \, C_s
\, {\rm Tr} \log \left( C^\dagger C + M_s^2 \right)
$$
\begin{equation}
\label{pauli2}+ \,\frac{1}{2} \sum_{s=0}^1 \, C_s \, {\rm Tr} \log \left(
C_0^\dagger C_0 + M_s^2 \right) \;,
\end{equation}
where $C_0 = 1$, $C_1 = -1$, $M_0 = 0$ and $M_1 = M$. This form makes it
explicit the fact that the finite-$M$ overlap in two spacetime dimensions
yields a regularized real part of the effective action, since it corresponds
to a Pauli-Villars like regularization, with $M$ playing the role of the
regulator mass. Moreover, this regularization is gauge invariant, since it
is provided by an operator depending on $C^\dagger C$, which transforms in a
gauge covariant way. It also follows from (\ref{pauli2}) that the real part
of the overlap effective action is {\em not regularized\/} in more than two
dimensions, since more Pauli-Villars regulators would be required.

Proceeding analogously to the case of the real part, it is possible to write
the imaginary part of the overlap effective action as the functional trace
(over the same functional space) of an operator:
\begin{equation}
\label{mfin}{\rm Im}\;\Gamma _{{\cal O}}\;=\;\frac 1{2i}Tr\left[ \log \left(
\frac{{\cal C^{\dagger }}{\cal C}_0}{(1+{\cal C^{\dagger }}{\cal C}_0)^2}%
\right) -\log \left( \frac{{\cal C}_0^{\dagger }{\cal C}}{(1+{\cal C}%
_0^{\dagger }{\cal C})^2}\right) \right] \,=\,\frac 1{2i}(A-A^{*})
\end{equation}
where
\begin{eqnarray}
A&\,=\,& Tr[ \log( {\cal C}^{\dagger} {\cal C}_0 (1+ {\cal C}^{\dagger}
{\cal C}_0)^{-2}]
\qquad  \nonumber\\
{\cal{C}} &\,=\,& C \, ({\hat \omega}+M)^{-1} \qquad
{\hat \omega}\,=\, (C^{\dagger} C + M^2)^{1\over2} \nonumber\\
{\cal C}_0 &\,=\,& C_0 \,({\hat \omega_0}+M)^{-1} \qquad
{\hat \omega_0}\,=\, (C_0^{\dagger} C_0 + M^2)^{1\over2} \;.
\label{ext}
\end{eqnarray}
As a check, one can consider the formal limit $M \to \infty$
\begin{equation}
\label{ima}{\rm Im} \; \Gamma_{{\cal O}}{\stackrel{M\to\infty}{=}} \frac{1}{%
2i} Tr\left[ \log \left(\frac{ C^{\dagger} C_0}{4 M^2}\right) -\log \left(%
\frac{C_{0}^{\dagger} C}{4 M^2}\right)\right] \;,
\end{equation}
which, at the unregularized level, is a sensible definition of the imaginary
part of the effective action for the corresponding chiral determinant. It is
evident from (\ref{mfin}) that ${\rm Im} \Gamma_{{\cal O}}$ has the
following structure:
\begin{equation}
\label{phi}{\rm Im} \, \Gamma_{{\cal O}} \;=\; {\rm Im} \, {\rm Tr} F ({\cal %
C}^\dagger {\cal C}_0) \;,
\end{equation}
with $F(x) = \displaystyle{\frac{x}{(1+x)^2}}$. This functional form of $F$
is not accidental, as the following argument shows: From the definition (\ref
{over}) of the overlap it follows that ${\rm Im} \Gamma$ is odd in $M$,
\begin{equation}
\label{odd}{\rm Im} \Gamma_{{\cal O}} (A,-M) \;=\; - {\rm Im} \Gamma_{{\cal O%
}} (A,M) \;.
\end{equation}
When this condition is impossed on (\ref{phi}) (assuming $F$ real), it
requires for $F$ to satisfy the functional equation
\begin{equation}
\label{funct}F (\frac{1}{x}) \;=\; F (x)
\end{equation}
condition which is satisfied by ${\displaystyle F(x) = \frac{x}{(1+x)^2}}$,
but not by an arbitrary function. This very same property is indeed imcompatible with a
Pauli-Villars like regularization of the imaginary part, since, as it will
be shown in the next section, that regularization would require, for
example, in two dimensions, a functional form
\begin{equation}
F(x) \;=\; \frac{x}{1+x} \;,
\end{equation}
which fails to satisfy (\ref{funct}). We have concluded that condition (\ref
{odd}) is incompatible with a regularized imaginary part of the overlap's
effective action, at least when a Pauli-Villars like regularization is used
(which is the case for the consistent regularization).


\section{Anomalous Divergence of the Chiral Current.}

The chiral anomaly can also be derived by taking the covariant divergence of
the expectation value $j_\mu^a$ of the chiral current operator $J_\mu^a$. We
shall now derive the anomaly following this procedure for the unregularised
overlap.

To start with, we recall the definition
\begin{equation}
\label{covdef}j_\mu^a \;=\; \langle J_\mu^a \rangle \;=\; {\rm tr} \left[
\gamma_\mu \tau_a S_F (x , y) \right]
\end{equation}
where $S_F (x , y)$ is the propagator for a chiral fermionic field in the
presence of an external gauge field. To evaluate (\ref{covdef}) in terms of
objects defined in the overlap formalism, we just need the overlap
definition for the chiral fermion propagator~\cite{daemi2}
$$
S(x,y) \;=\; \lim_{M \to \infty} S(x,y; M)
$$
\begin{equation}
S(x,y; M) \;=\;\frac{1}{M} \, \frac{\langle A + | \Psi (x) \Psi^\dagger (y)
| A - \rangle}{\langle A + | A - \rangle} \;.
\end{equation}

By using the definitions presented in section 2, we obtain, after some
algebra,
\begin{equation}
S(x,y;M) \;=\; \frac{1}{2 M} \, \langle x |\left(
\begin{array}{cc}
1 & -
\displaystyle{\frac{1}{{\hat \omega} - M}} C^\dagger \\ - C \displaystyle{%
\frac{1}{{\hat \omega} + M}} & 1
\end{array}
\right)|y \rangle \;,
\end{equation}
expression which is valid even if zero modes are present.

We then take the covariant derivative of the current calculated with the
finite-$M$ propagator
$$
(D_\mu j_\mu )_a (x) \;=\; \frac{1}{2M}\, \lim_{y \to x} \, {\rm tr} \left[
\gamma_\mu \tau_a D_\mu^x S(x,y;M) \right]
$$
$$
=\,\frac{1}{2M} \lim_{y \to x} {\rm tr} \left\{ \langle x | \tau_a \left(
\begin{array}{cc}
0 & - C^\dagger \\
C & 0
\end{array}
\right) \left(
\begin{array}{cc}
1 & -
\displaystyle{\frac{1}{{\hat \omega} - M}} C^\dagger \\ - C \displaystyle{%
\frac{1}{{\hat \omega} + M}} & 1
\end{array}
\right) |y \rangle \right\}
$$
\begin{equation}
\label{andiv}= \; \frac{1}{2 M} \, \lim_{y \to x} \left[ \langle x | {\rm tr}
(\tau_a {\hat \omega} ) | y \rangle - \langle x | {\rm tr} ( \tau_a {\tilde
\omega} ) | y \rangle \right] \;,
\end{equation}
where we have defined
\begin{equation}
{\tilde \omega} \;=\; ( C C^\dagger \,+\, M^2 )^{\frac{1}{2}} \;.
\end{equation}
We note that a formula that resembles Fujikawa's expression for the anomaly
can be derived by simply rewriting (\ref{andiv}) as
\begin{equation}
\label{fuji}(D_\mu j_\mu)_a (x) \;=\; \frac{1}{2} \, \frac{M}{|M|} \, \int
\frac{d^d k}{(2 \pi)^d} \, e^{-i k \cdot x} {\rm tr} [ \gamma_5 \tau_a f(-%
\frac{\not \! D^2}{M^2}) ] e^{i k \cdot x}
\end{equation}
with $f (x) \;=\; \sqrt{ 1 \, + \, x }$. It is evident that the result it
yields for the anomaly is unregularized, since the would-be regulating
function $f$ in this case fails to satisfy the condition $f(\infty)=0$,
necessary to make (\ref{fuji}) convergent. However, it is worth remarking
that the integral over $x$ of (\ref{andiv}) is meaningful, since
\begin{equation}
\int \, d^d x \, D_\mu j_\mu (x) \;=\; \frac{1}{2} \frac{M}{|M|} \, \left(
n_+ \, - \, n_- \right)
\end{equation}
where $n_+$ ($n_-$) is the number of zero-modes of the operator $C$ ($%
C^\dagger$). Hence, the integral of the anomaly is proportional to the index
of the operator $C$, as it should be.

This calculation also illustrates what we said at the end of the previous
section, since the anomaly (proportional to the variation of the imaginary
part) is odd in $M$ and divergent.

\section{Regularization.}

The real part ${\rm Re} \Gamma_{{\cal O}}$ may be regularized in a
gauge-invariant way by introducing in (\ref{real}) a regulating function
depending on the operator $C^\dagger C$, which transforms in a
gauge-covariant way:
$$
{\rm Re} \; \Gamma_{{\cal O}}(A,M,\Lambda) \,=\, - \displaystyle{\frac{1}{2}}
{\rm Tr} \log \left[ f\left(\frac{C^\dagger C}{\Lambda^2}\right) \frac{%
C^\dagger C}{C^\dagger C +M^2}\right]
$$
\begin{equation}
\label{rereg}+\;\displaystyle{\frac{1}{2}} {\rm Tr} \log \left[ f\left(\frac{%
C_0^\dagger C_0}{\Lambda^2}\right) \frac{C_0^\dagger C_0}{C_0^\dagger C_0
+M^2} \right] \;,
\end{equation}
where $f$ satisfies the conditions
\begin{equation}
\label{condf}f(z)\,{\stackrel{z\to 0}{=}}\,1 \;\;\; ; f(z)\,{\stackrel{z\to
\infty}{=}}\,f^{\prime}(z)\,{\stackrel{z\to \infty}{=}}\, \cdots f^{(n)}(z)\,%
{\stackrel{z\to \infty}{=}}\,0\;\;.
\end{equation}
For finite $\Lambda$, the $M \to \infty$ limit may be taken safely in (\ref
{rereg}), since conditions (\ref{condf}) assure the convergence of the
functional trace.

To regulate the imaginary part is not so straightforward as for the real
part. Simple algebraic manipulations in (\ref{mfin}) lead to the following
convenient expression for the imaginary part,
\begin{equation}
\label{14}{\rm Im}\Gamma _{{\cal O}}(A,M)\,=\,{\rm Im}\left\{ {\rm Tr}[\log
(W_0\;I\;W)]\right\} \;,
\end{equation}
where,
\begin{eqnarray}
W_0 &\,=\,& {\hat \omega}_0 + M \nonumber\\
W &\,=\,& {\hat \omega} + M \nonumber\\
I &\,=\,& (W W_0 + C^{\dagger} C_0 )^{-1} \, C^{\dagger} C_0
(W W_0 + C^{\dagger} C_0 )^{-1}  \; .
\label{15}
\end{eqnarray}
It is worth noting that if the property,
\begin{equation}
\label{16}{\rm Tr}[\log (ab)]\,=\,{\rm Tr}[\log (a)+\log (b)]\;,
\end{equation}
were valid for the factors in (\ref{14}) then the contribution from the two
positive-definite hermitean factors $({\hat \omega }_0+M)$ and $({\hat
\omega }+M)$ should cancel in (\ref{14}) when taking the imaginary part.
Regularizing these factors by,
\begin{eqnarray}
W_{0}^{reg} &\,=\,& e^{-W_{0}/\Lambda} \, W_{0}
\nonumber\\
W^{reg} &\,=\,& e^{-W/\Lambda} \, W \;,
\label{17}
\end{eqnarray}
(\ref{16}) holds~\footnote{ Using the Backer-Haussdorff relation
it is simple to show that (\ref{16}) is valid if,
$$
{\rm Tr}([\log a,\log b] + {\rm additional \; nested \; commutators})
\,=\,
0 \;\,
$$
by means of theorem VI.25 of Ref.(\cite{rs}) this is assured if one of the
factors of the commutators is trace class and the other is bounded.
This is so when employing the regularized operators (\ref{17}) in this
first regularized version ${\rm Im} {\bar \Gamma}_{\cal O} (A,M)$ of
(\ref{14}).}.
Hence we have,
\beq
{\rm Im} {\bar \Gamma}_{\cal O} (A,M) \,=\,
{\rm Im } \left\{ \rm Tr \left[
\log I \right] \right\} \;.
\label{18}
\eeq
The left hand side of (\ref{18}) has not yet been regularized. This
task is rendered difficult since the non-hermitean operator
$C^{\dagger} C_0$ is not even normal. Hence we can not assert whether it has
a well defined eigenvalue problem or not. In this work we will only look for a
perturbatively well-defined version of (\ref{18}). It turns out that the
following regularization works in the above sense,
\bea
{\rm Im} {\bar
\Gamma}^{reg}_{\cal O} (A,M) &\,=\,&
{\rm Im} \left\{
\rm Tr \left[ \log
\left( \varphi ( \frac{C^\dagger C_0}{\Lambda^2} ) (W W_0 + C^{\dagger}
C_0 )^{-1}
\right. \right. \right. \,\nonumber\\
& & \left. \left. \left.
C^{\dagger}
C_0 (W W_0 + C^{\dagger} C_0 )^{-1} \right) \right]
\right\}   \;,
\label{19}
\eea
where $\varphi$ satisfies conditions analogous to (\ref{condf}) for $f$, and
its explicit form depends on the spacetime dimension. It is convenient to pass from
(\ref{19}) to a simplified version which however yields the same results when
$M$ is taken to be infinite. The simplified formula follows from the
observation that, for a finite $\Lambda$, one can take $M$ to be
much larger than $\Lambda$ and thus than all the momentum scales
in the problem. It is then possible to make a large-$M$ expansion,
which to leading order amounts to:
\beq
{\hat \omega} \; \simeq \; {\hat \omega_0} \; \simeq \; |M|
\;\;\; , \;\;\; W W_0 \; \simeq \; 4 M^2 \;.
\eeq
Thus the regularized imaginary part is approximated by
\bea
{\rm Im} {\bar
\Gamma}^{reg}_{\cal O} (A,M) &=&
{\rm Im} \left\{
\rm Tr \left[ \log
\left( \varphi (\frac{C^{\dagger} C_0}{\Lambda^2}) (4 M^2 + C^{\dagger}
C_0 )^{-1}
\right. \right. \right. \,\nonumber\\
& & \left. \left. \left.
C^{\dagger}
C_0 (4 M^2 + C^{\dagger} C_0 )^{-1} \right) \right]
\right\}   \;.
\label{imr}
\eea
Incidentally, this expression is tantamount to making the replacements
${\cal C} \to C / (2 M) $, ${\cal C}_0 \to C_0 /(2 M)$ in (\ref{19}).
This can be understood from the fact that in a polar decomposition both
${\cal C}$ and $C$ have the same unitary part. This part should be the only
important one in the determination of the imaginary part at large $M$.
We have also explicitly verified that an expansion in powers of $A_\mu$
of the finite-$M$ expression (\ref{18}) has the same  superficial
degree of divergence as the one approximated for large values of $M$.

An unpleasant feature of expression (\ref{imr}) is that it depends
on the two masses $M$ and $\Lambda$, which should eventually tend
to infinity. This introduces the problem of deciding an
order for taking those limits. We discard the order where $\Lambda$ is taken
to infinity before $M$, since we know that we then get an expression
where $\varphi$ is replaced by one, which is not regularized. The natural
order seems to be the one where $M$ is taken to infinity first,
with the proviso that the form of $\varphi$ should be chosen in order
to render the imaginary part finite even when $M \to \infty$
(what can always be done). When this limit has been taken, we are
led to an expression of the form
\beq
{\rm Im} {\bar \Gamma}^{reg}_{\cal O} (A) \;=\;
{\rm Im} \left\{ \rm Tr \left[ \log
\left( \varphi (\frac{C^{\dagger} C_0}{\Lambda^2})
\frac{C^{\dagger} C_0}{\Lambda^2} \right) \right] \right\}   \;,
\label{imr1}
\eeq
where the factor $\frac{1}{\Lambda^2}$ has been introduced to render the
argument of the logarithm dimensionless (it does not affect the imaginary
part, though).

Next we apply the previously
discussed concepts to the two-dimensional
non-Abelian case in perturbation
theory.
It is straightforward to see that the real part is Pauli-Villars
regulated
for a finite $M$. The perturbative series for
${\rm Re} \;
\Gamma_{\cal O}$ can be put as
\beq
{\rm Re} {\Gamma}_{\cal O} \;=\; -
\frac{1}{2} \, \sum_{r=1}^\infty \,
\frac{(-1)^{r-1}}{r} \, \sum_{s=0}^1 \,
C_s \, {\rm Tr} \,
\left\{ [ (-\partial^2 + M_s^2)^{-1} V ]^r \right\}
\label{pertex}
\eeq
where
\beq
V \;=\; C^\dagger_0 \sigma \cdot A \,-\,
\sigma^\dagger \cdot A C_0 \,-\,
\sigma^\dagger \cdot A \, \sigma \cdot
A
\eeq
and we have used the property $C^\dagger_0 C_0 = - \partial^2$. To
realize that (\ref{pertex}) is regulated, we just note that the
superficial degree of divergence of the ($s=0$ or $s=1$) term of
order $r$
is $\omega_r = 2 - r$. Hence all the terms with $r > 2$
converge (even for
an infinite $M$). We only need to consider
the terms with $r=1$ and $r=2$,
which shall be denoted by
$\gamma_1$ and $\gamma_2$, respectively. $%
\gamma_1$ may
be written as
\beq
\gamma_1 (A,M) \;=\; - \frac{1}{2} \int
\frac{d^2 k}{(2 \pi)^2}
\frac{M^2}{k^2 (k^2 + M^2)} \sigma^\dagger_\mu
\sigma_\nu
\int d^2 x \, A^a_\mu (x) A^a_\mu (x) \;,
\eeq
which is
obviously convergent. The second order term $\gamma_2$ is
given by
$$
\gamma_2 \;=\; \frac{1}{4} \, \sum_{s=0}^1 C_s {\rm Tr}
\left\{ [
(-\partial^2 + M^2)^{-1} (C_0^\dagger \sigma \cdot A -
\sigma^\dagger \cdot A C_0 )(-\partial^2 + M^2)^{-1} \right.%
$$
\beq
\left. (C_0^\dagger \sigma \cdot A - \sigma^\dagger \cdot C_0 ) ]
+
[(-\partial^2 + M^2)^{-1} \sigma^\dagger \cdot A  \sigma \cdot A
(-\partial^2 + M^2)^{-1}\sigma^\dagger \cdot A ] \right\}\;.
\eeq
To
make the degree of divergence explicit, we take the trace in
momentum
space. One sees that it contains the two momentum integrals
$$
I_{1 \mu\nu} \;=\; \int \frac{d^2 k}{(2 \pi)^2}
\sum_{s=0}^1 C_s \left\{
\frac{k_\mu (k+p)_\nu}{(k^2 + M_s^2)[(k+p)^2 + M_s^2]}
\right\} $$

\beq
I_2 \;=\; \int \frac{d^2 k}{(2 \pi)^2} \sum_{s=0}^1 C_s
\left\{
\frac{1}{(k^2 + M_s^2)[(k+p)^2 + M_s^2]}
\right\} \;.
\eeq
A simple
analysis of the large momentum behaviour of these integrals
shows that they
are convergent, with superficial degrees of divergence
$\omega = -2$ and $%
\omega = -4$ for $I_1$ and $I_2$, respectively.
This completes the proof of
the finiteness of the real part for the
two dimensional case.

Regarding the imaginary part, the form of the regulating function which
is used to make it finite can always be chosen in a Pauli-Villars like
fashion, which corresponds to the consistent anomaly~\cite{ball}.
It is worth noting, however, that if one accepts a definition of the
imaginary part where the term $A$ defined in (\ref{ext}) is divergent,
but the divergencies cancell when taking its imaginary part, no
regularization is needed, at least in two dimensions. This, however, is
a very special way of evaluating the imaginary part, and one can say
that amounts to a kind of 'symmetric' regularization. For the Abelian
case in two dimensions, this evaluation yields, after taking the limit
$M \to \infty$:
\beq
{\rm Im} {\bar \Gamma}_{\cal O} (A)
\,=\,  \frac{1}{4 \pi} \, \int d^2 x \, \partial \cdot A \,
\frac{1}{\partial^2} \, \epsilon_{\mu \nu} \partial_\mu A_\nu \;,
\eeq
which yields the right value for the anomaly when taking the gauge
variation.

The results about regularization are summarized in the table bellow.
They depend on the dimension and on whether we are considering the real or
the imaginary part.
\beq
\begin{array}{ccc}
\;  & {\rm Real} & {\rm
Imaginary} \\
d=2 & {\rm Converges\; with}f=1 & {\rm Converges \; with} f=1
\\
    & {\rm Regularization \; not\; necessary.}
    &{\rm However
\;neither}\; A \;{\rm nor} \;A^{\dagger} {\rm \;converge.}\\
d=4 & {\rm
\;Necessary. Converges \;for} & {\rm Necessary. Converges \;at} \\
    &
{\rm adequate \;choice \;of} f.
    &{\rm least \;perturbatively \;for
\;adequate} \, f.
\end{array}
\eeq
%
\section{Anomalies}
We start from expression (\ref{imr}). Its gauge
variation is given by,
\beq
\delta
{\rm Im} {\bar \Gamma}^{reg}_{\cal O} (A)
\,=\, - \frac{1}{2i} \left\{
Tr \left[ \varphi\left(%
\frac{C_0^{\dagger} C}{\Lambda^2}\right)
\frac{C_0^{\dagger} \delta C}{%
\Lambda^2}\right]
- Tr \left[ \varphi\left(\frac{C_0 C^{\dagger}}{\Lambda^2}%
\right)
\frac{C_0 \delta C^{\dagger}}{\Lambda^2} \right] \right\} \;,
\label{ca}
\eeq
where,
\beq
\varphi(x) \,=\, f^{-1}(x) \frac{df(x)}{dx}
+ \frac{1}{x} \;,
\label{de}
\eeq
rewriting this last expression in Dirac
space we are lead to,
\beq
\delta
{\rm Im} {\bar \Gamma}^{reg}_{\cal O} (A)
\,=\,\frac{1}{2i\Lambda^2}
Tr\left\{ \varphi\left(- \frac{\not \!\partial \not  \!\! D}{%
\Lambda^2}\right)
\gamma_5 \not \!\partial [\not  \!\! D,\omega]\right\}
\;,
\label{d}
\eeq
$\omega$ being, as before, the infinitesimal parameter
of the gauge transformation. Defining ${\cal A}_a (x)$ by,
\beq
\delta
\phi_{reg}(A,\Lambda) \,=\,
\int d^d x \; \omega_a(x) {\cal A}_a (x)
\;,
\eeq
we get,
\beq
{\cal A}_a (x) \,=\, D_{\mu}^{ab} J_{\mu}^b (x)
\;,
\eeq
with,
\beq
J_{\mu}^a (x) \,=\, - \,
\frac{1}{2i\Lambda^2}
{\rm Tr}
\langle x | \varphi \left(- \frac{ \not \!\partial \not  \!\! D}
{\Lambda^2} \right)
\gamma_5 \not \! \partial  \gamma_{\mu}
\tau^a | x \rangle \;,
\eeq
for $\Lambda^2 \to \infty$ we obtain the
unregulated expression,
\beq
J_{\mu}^a (x) \,=\, - \,
\frac{1}{2i}
Tr \langle
x |- \frac{1}{\not \!\partial \not  \!\! D}
\gamma_5 \not \! \partial
\gamma_{\mu} \tau^a|x\rangle \;,
\eeq
showing that (\ref{d}) corresponds
to a regularization of the
first order variation of the chiral determinant,
therefore
leading to a consistent anomaly.

We conclude that,
\begin{%
enumerate}
\item[(i)] When regularized the overlap claim (\ref{oc}) is
correct.
\item[(ii)] The parameter $M$ can be interpreted as a
regularization
parameter only for the real part of the overlap effective
action
($\Gamma_{\cal O}$) in $d=2$.
\item[(iii)]The regularized overlap
leads, as it should be, to consistent
anomalies for the gauge variation of $%
\Gamma_{\cal O}$.
\end{enumerate}

\section*{Acknowledgements.}

We acknowledge conversations with Profs. R. Iengo and S. Randjabar-Daemi.
C. D. F. thanks Fundaci\'on Antorchas for financial support. R. C. T.
does the same with the ICTP.
\newpage
\begin{%
thebibliography}{99}
\bibitem{ball}For an excellent review see, for
example,
R. D. Ball, Phys. Rep. {\bf 182}, numbers 1 and 2 (1989).
\bibitem{kap}D. B. Kaplan, Phys. Lett. {\bf B288} 342 (1992).
\bibitem{slav1%
}S.A. Frolov and A.A. Slavnov, Phys.Lett.
{\bf B309} 344 (1993).
\bibitem{%
slav2}A.A. Slavnov, Phys.Lett. {\bf B348} 553 (1995).
\bibitem{toft} G.
t'Hooft.
\bibitem{nar}R. Narayanan and H. Neuberger, Nucl.
Phys. {\bf
B443}:305-385 (1995). See also:
R. Narayanan, H. Neuberger and P. Vranas,
hep-lat/9509046;
R. Narayanan and H. Neuberger, hep-lat/9509047;
H.
Neuberger, \\ hep-lat/9511001.
\bibitem{daemi1}S. Randjbar-Daemi and J.
Strathdee,
Phys. Lett. {\bf B348}, 543 (1995).
\bibitem{daemi2}S.
Randjbar-Daemi and J. Strathdee,
Nucl. Phys. {\bf B443} 386 (1995)
\bibitem{daemi3}S. Randjbar-Daemi and  J. Strathdee,
Phys. Rev. {\bf D51},
6617 (1995).
\bibitem{daemi4} S. Randjbar-Daemi and J. Strathdee,
Nucl.
Phys. {\bf B466}, 335 (1996).
\bibitem{rs} M. Reed and B. Simon , {\bf
Functional Analysis},
Academic Press, 1972.
\bibitem{cesar1} C.D. Fosco,
Int. J. Mod. Phys. {\bf A11} 3987 (1996).
\end{thebibliography}

\end{document}